\documentclass[aps,prd,amsmath,twocolumn]{revtex4}

\usepackage{amsmath}
\usepackage{bm}
\usepackage[dvips]{color,graphicx}
\usepackage{epsfig}
\usepackage{dcolumn}
\newcommand{\sw}{\sin^2\theta_W}
\usepackage{latexsym}
\def\slash#1{{\rlap{\hspace{.08em}/}#1}}

\begin{document}

\title{$\gamma Z$ corrections to forward-angle parity-violating
        $e\, p$ scattering}

\author{A. Sibirtsev$^{1,2}$,
	P. G. Blunden$^{3,2}$,
	W. Melnitchouk$^2$ and
	A. W. Thomas$^4$}
\affiliation{
$^1$Helmholtz-Institut f\"ur Strahlen- und Kernphysik (Theorie), 
    Universit\"at Bonn, 
    D-53115 Bonn, Germany				\\
$^2$Jefferson Lab, 12000 Jefferson Avenue, Newport News,
    Virginia 23606, USA\\
$^3$Department of Physics and Astronomy,
    University of Manitoba, Winnipeg, MB, Canada R3T 2N2\\
$^4$CSSM, School of Chemistry and Physics, University of Adelaide,
    Adelaide SA 5005, Australia}

\begin{abstract}
We use dispersion relations to evaluate the $\gamma Z$ box contribution
to parity-violating electron scattering in the forward limit arising
from the axial-vector coupling at the electron vertex. The calculation
makes full use of the critical constraints from recent JLab data on
electroproduction in the resonance region as well as high energy data
from HERA. At the kinematics of the ${\rm Q_{weak}}$ experiment,
this gives a correction of $0.0047^{+0.0011}_{-0.0004}$ to the
Standard Model value $0.0713(8)$ of the proton weak charge.
While the magnitude of the correction is highly significant, the
uncertainty is within the anticipated experimental uncertainty of
$\pm 0.003$.
\end{abstract}

\maketitle

Amongst the many methods for searching for physics beyond the
Standard Model, the verification of the predicted evolution of the
Weinberg angle from the $Z$-pole to very low energies is currently
of great interest.
In particular, the ${\rm Q_{weak}}$ experiment at Jefferson Lab
\cite{QWEAK} is designed to measure the weak charge of the proton
using parity-violating elastic electron scattering (PVES) from the
proton to a higher level of precision than previously possible.
In combination with constraints from atomic parity violation
\cite{AtomicPV}, ${\rm Q_{weak}}$ aims to either discover evidence
for new physics beyond the Standard Model that leads to parity
violation in electron scattering or raise the limit on its mass
scale to above 2~TeV, complementing direct searches at the LHC
\cite{YoungSM,Erler}.

In PVES the parity-violating asymmetry in the $t \to 0$ and low energy
limit is related to the weak charge of the proton $Q_W^p$ \cite{Musolf}:
\begin{eqnarray}
A^{\rm PV}
\equiv	\frac{\sigma_R - \sigma_L}{\sigma_R + \sigma_L}
\to \frac{G_F}{4 \pi \alpha \sqrt{2}}\, t\,Q_W^p\, ,
\label{eq:Apv}
\end{eqnarray}
where $\sigma_{L(R)}$ is the cross section for left- (right-) hand
polarized electrons, $G_F$ is the Fermi constant, and $\alpha$ is the
fine structure constant.
The arrow serves to remind that this relation is only realized when
radiative corrections are properly accounted for, in particular, any
residual dependence on the electron energy $E$ or the momentum transfer
squared $t$.
Including electroweak radiative corrections, the proton weak charge
is defined at zero energy and momentum transfer as \cite{Erler}
\begin{eqnarray}
Q_W^p &=& (1 + \Delta\rho + \Delta_e)
	  (1 - 4 \sw(0) + \Delta_e')			\nonumber\\
      & & +\, \Box_{WW} + \Box_{ZZ} + \Box_{\gamma Z}(0)\ ,
\label{eq:Qwp}
\end{eqnarray}
where $\sw(0) = 0.23867(16)$ is the weak mixing angle at zero momentum,
and the corrections $\Delta\rho$, $\Delta_e$ and $\Delta_e'$ are given
in \cite{Erler} and references therein.
The contributions $\Box_{WW}$ and $\Box_{ZZ}$ from the $WW$ and $ZZ$
box diagrams can be computed perturbatively, while the $\gamma Z$
interference correction $\Box_{\gamma Z}(E)$ in addition depends on
physics at low momentum scales \cite{Erler,Musolf,MS}.
The current best theoretical estimate from Ref.~\cite{Erler} is
$Q_W^p=0.0713(8)$.

In Eq.~(\ref{eq:Qwp}) we have explicitly introduced a dependence
of $\Box_{\gamma Z}(E)$ on the electron energy $E$.
The energy-dependence of the other radiative corrections
in Eq.~(\ref{eq:Qwp}) is not expected to be important at the
${\cal O}$(GeV) energies relevant for PVES.
The $Q_W^p$ extracted from $A^{\rm PV}$ in Eq.~(\ref{eq:Apv})
will then differ from $Q_W^p$ in Eq.~(\ref{eq:Qwp}) by an amount
$\Box_{\gamma Z}(E)-\Box_{\gamma Z}(0)$, which we refer to in what
follows as a correction to $Q_W^p$ at the particular kinematics of
the electron scattering experiment.
In general the $\gamma Z$ term has contributions from the vector
electron--axial vector hadron coupling of the $Z$ boson
($\Box_{\gamma Z}^A$) and from the axial vector electron--vector hadron
coupling of the $Z$ ($\Box_{\gamma Z}^V$),
$\Box_{\gamma Z}(E) = \Box_{\gamma Z}^A(E) + \Box_{\gamma Z}^V(E)$.

Given that the ${\rm Q_{weak}}$ experiment has a precision target of
4.2\% on $Q_W^p$ \cite{QWEAK}, if we are to draw meaningful conclusions 
in relation to the Standard Model it is crucial that all the radiative
corrections to PVES be under control at a level well below this target.
The first studies of the box corrections \cite{Erler,Musolf,MS}
suggested that they were indeed understood to the required precision,
with the uncertainty on the least constrained, $\Box_{\gamma Z}(0)$
term being 0.65\%.

In their seminal early work, Marciano \& Sirlin \cite{MS} computed
the $\Box_{\gamma Z}^A(0)$ correction, which is dominant in atomic
parity-violation experiments at very low electron energies.
This correction was further divided into a high-momentum contribution
to the loop integral, computed at the quark level, and a low-momentum
contribution, computed with the nucleon elastic intermediate state.
The entire uncertainty on the calculation was taken to arise from
the low-energy component \cite{Erler}.

In a stimulating new analysis, Gorchtein and Horowitz \cite{GH} used
forward angle dispersion relations to evaluate the additional correction,
$\Box_{\gamma Z}^V(E)$, which is negligible at the low electron energies
characteristic of atomic parity violation. However,
at the ${\cal O}$(GeV) energies relevant for PVES \cite{QWEAK}, this
correction was found to be large, with an uncertainty potentially capable
of jeopardizing the interpretation of the ${\rm Q_{weak}}$ experiment.
Recent model-dependent analyses of the low-energy $\gamma Z$
contribution, involving only nucleon and $\Delta$ intermediate states,
find smaller but non-negligible effects
\cite{Elastic1,Elastic2,Elastic3}.

In this work we revisit this new $\Box_{\gamma Z}^V(E)$ radiative
correction with a detailed evaluation of the inelastic contributions,
taking full advantage of the wealth of data available in the resonance
region and from deep-inelastic scattering (DIS) at high energy.
We find that with these new experimental constraints the total
$\Box_{\gamma Z}^V(E)$ contribution is quite well under control,
constituting a correction of $0.0047^{+0.0011}_{-0.0004}$, or
$6.6^{+1.5}_{-0.6}\, \%$, to
$Q_W^p$ at the ${\rm Q_{weak}}$ kinematics.

The ${\rm Q_{weak}}$ experiment is designed to measure the PVES asymmetry
from protons at $E=1.165$~GeV electron energy and very low momentum
transfer squared, $-t = 0.026$~GeV$^2$ \cite{QWEAK}.
The purely electromagnetic and weak contributions are independent of the 
electron polarization, so that the asymmetry directly measures the 
parity-violating interference between the photon- and $Z$-exchange 
amplitudes. 

For the $e(k)+N(p)$ elastic scattering process in the forward angle 
limit, the Born amplitudes
\begin{eqnarray}
{\cal M}_\gamma
&=& {1 \over t} \left( -e\, 2 k_\mu \right)
		\left( +e\, 2 p^\mu \right)\
 =\ -{4\pi\alpha \over t} 4 k\cdot p\, ,	\\
{\cal M}_Z^{\rm PV}
&=& \left( {-2 G_F\over\sqrt{2}} \right)
    \left( -g_A^e 2 k_\mu \right)
    \left( \frac{1}{2} Q_W^p 2 p^\mu \right)	\nonumber\\
&=& {G_F\over \sqrt{2}} g_A^e Q_W^p 4 k\cdot p\, ,
\end{eqnarray}
depend only on the electron and proton convection currents.
The PV $Z$ exchange amplitude ${\cal M}_Z^{\rm PV}$ incorporates
the difference between right- and left-polarized electron currents,
so that
$A^{\rm PV} = 2 {\cal M}_\gamma^* {\cal M}_Z^{\rm PV}/|{\cal M}_\gamma|^2$
in Eq. (\ref{eq:Apv}) with $g_A^e=-1/2$ the weak axial charge of
the electron.

The correction $\Box_{\gamma Z}$ in Eq.~(\ref{eq:Qwp})
arises from the replacement
${\cal M}_Z^{\rm PV} \to
 {\cal M}_Z^{\rm PV}+{\cal M}_{\gamma Z}^{\rm PV}$,
so that
\begin{equation}
\Box_{\gamma Z}
= Q_W^p {{\cal M}_\gamma^* {\cal M}_{\gamma Z}^{\rm PV}
	\over{\cal M}_\gamma^* {\cal M}_Z^{\rm PV}}\, .
\end{equation}
Corrections from the interference of ${\cal M}_Z^{\rm PV}$ with
the two-photon exchange amplitude under the replacement
${\cal M}_\gamma \to {\cal M}_\gamma+{\cal M}_{\gamma\gamma}$
vanish in the $t\to 0$ limit, and therefore do not affect the asymmetry.

Applying Cauchy's integral theorem, the real part of
$\Box_{\gamma Z}^V(E)$ can be obtained from its imaginary part
using a standard dispersion relation,
\begin{equation}
\Re e\, \Box_{\gamma Z}^V(E)
= {1 \over \pi}
  P \int_{-\infty}^\infty dE'
  {\Im m\, \Box_{\gamma Z}^V(E') \over E'-E}\ ,
\label{eq:DR}
\end{equation}
where $P$ is the principal value.
The integration over negative energies corresponds to the crossed
$\gamma Z$ box diagram, with the vector hadron correction even under
$E' \to -E'$.
Consequently $\Re e\, \Box_{\gamma Z}^V(0)=0$, justifying the neglect
of this term at atomic scale electron energies.
Following Ref.~\cite{GH} we compute only the inelastic contributions
to the dispersion integral; the elastic component has previously been
computed to be small \cite{MS,Elastic1,Elastic2,Elastic3}.

From the optical theorem, the imaginary part of PV $\gamma Z$ exchange 
amplitude ${\cal M}_{\gamma Z}^{\rm PV}$ can be written in terms of the 
cross section for all possible final hadronic states
\begin{eqnarray}
2\, \Im m\, {\cal M}_{\gamma Z}^{\rm PV}
&=& 4\pi M \int {d^3 k'\over (2\pi)^3 2 E_{k'}}
    \left({4\pi \alpha\over Q^2}\right)
    \left( {-2 G_F\over\sqrt{2}}\right) \nonumber\\
& & \times {1\over 1+Q^2/M_Z^2} L_{\mu\nu}^{\gamma Z}
				W^{\mu\nu}_{\gamma Z}\, ,
\end{eqnarray}
where $q=k-k'$ is the virtual four-momentum transfer (with $Q^2=-q^2$),
$M (M_Z)$ is the proton ($Z$-boson) mass, and (neglecting the mass of 
the electron) 
$L_{\mu\nu}^{\gamma Z}
= \frac{1}{2}{\rm Tr}\left[\gamma_\mu\slash{k}'\gamma_\nu
  (g_V^e-g_A^e\gamma_5)\gamma_5 \slash{k}\right]$
is the PV leptonic tensor arising from the difference between right
and left polarized electrons.
The symmetric part of the hadronic tensor
\begin{eqnarray}
W^{\mu\nu}_{\gamma Z}
&=& {1\over 4\pi M} \int d^4 x\,e^{i q\cdot x}	\nonumber\\
& & \times
\langle p \left| 
\left[ J_\gamma^\mu(x) J_Z^\nu(0) + J_Z^\mu(x) J_\gamma^\nu(0) \right]
\right| p\rangle
\end{eqnarray}
can be written in terms of the interference electroweak 
$F_{1,2}^{\gamma Z}$ structure functions,
\begin{equation}
W^{\mu\nu}_{\gamma Z}
= {1\over M}
  \left[- F_1^{\gamma Z} g^{\mu\nu}
	+ F_2^{\gamma Z} {p^\mu p^\nu\over p\cdot q}
  \right] .
\end{equation}
Making a change of variables
\begin{equation}
{d^3 k'\over (2\pi)^3 2 E_{k'}}
\to {1\over 32 \pi^2 k\cdot p} dW^2\, dQ^2\, ,
\end{equation}
and evaluating $k\cdot p = M E$ in the rest frame of the proton,
we find for the $g_A^e$-dependent part:
\begin{eqnarray}
\Im m\, \Box_{\gamma Z}^V(E)
&=&
\frac{\alpha}{(s - M^2)^2} 
\int\limits_{W_\pi^2}^s\!\! dW^2 \!\!\!
\int\limits_0^{Q^2_{\rm max}} \!\!\!
\frac{dQ^2}{1+Q^2/M_Z^2}  			\nonumber \\
& & \hspace*{-1.5cm}
\times
\left[ F_1^{\gamma Z}
     + F_2^{\gamma Z}{s \left(Q^2_{\rm max}-Q^2\right) \over 
       Q^2 \left(W^2 - M^2 + Q^2\right)}
\right] ,
\label{eq:imagi}
\end{eqnarray}
where $s = M (M+2E)$ is the total c.m.\ energy squared.
The structure functions $F_{1,2}^{\gamma Z}$ are functions of the
exchanged boson virtuality, $Q^2$, and of the invariant mass $W$
of the exchanged boson and proton (or alternatively of the Bjorken
variable $x = Q^2/(W^2-M^2+Q^2)$).
The lower limit of the $W$ integration is given by the mass
$W_\pi = M + m_\pi$ of the pion production threshold, and the upper limit
of the $Q^2$ integration is given by $Q^2_{\rm max} = 2ME (1-W^2/s)$.
Our definitions of the structure functions coincide with the standard
definitions given by the PDG \cite{PDG}, in which at large $Q^2$ and $W$
the $F_2^{\gamma Z}$ structure function, for example, can be written
(at leading order) in terms of parton distributions $q$ and $\bar q$ as
\begin{eqnarray}
F_2^{\gamma Z} &=& x \sum_q 2 e_q g_V^q\, ( q + \bar q )\, ,
\end{eqnarray}
with weak vector charges $g_V^u = 1/2 - (4/3) \sw$ and
$g_V^d = -1/2 + (2/3) \sw$ for $u$ and $d$ quarks, respectively.
In particular, in the limit $2 g_V^q \to e_q$ the interference
structure functions $F_{2}^{\gamma Z} \to F_{2}^\gamma$ \cite{PDG},
where
\begin{eqnarray}
F_2^\gamma  &=& x \sum_q e_q^2\, ( q + \bar q )\, .
\end{eqnarray}

We note that the expression (\ref{eq:imagi}) is a factor of 2 larger
than that quoted in Ref.~\cite{GH}. Because of the importance of this
difference we have carried out a number of checks of our result.
Most importantly, we have verified that Eq.~(\ref{eq:imagi}) reproduces
the known asymptotic limit for a point-like hadron \cite{MS}, as well as
the independently calculated result for an elastic nucleon intermediate
state \cite{Elastic1,Elastic2,Elastic3}.
We also point out that the relation between the structure functions and
the virtual photon total cross sections used in \cite{GH} omits a factor
$(1-x)$ relative to the usual definition, which leads to an overestimate
of the contribution by 30--40\% (and even more in the resonance region).

In the region of low intermediate state hadronic masses,
$W \lesssim 2.5$~GeV, inclusive scattering is dominated by nucleon
resonances. While there is an abundance of electroproduction data
in the resonance region, there are no direct measurements of
$F_{1,2}^{\gamma Z}$.
For transitions to isospin $I=3/2$ states, such as the $\Delta$
resonance, CVC and isospin symmetry dictate that the weak isovector
transition form factors are equal to the electromagnetic ones
multiplied by $(1+Q_W^p)$.
For isospin $I=1/2$ resonances, which contain contributions from
isovector and isoscalar currents, using SU(6) quark model wave
functions one can verify that for the most prominent $I=1/2$ states
the magnitudes of the $Z$-boson transition couplings are equal to
the respective photon couplings to within a few percent.

\begin{figure}[tH]
\vspace*{-6mm}
\hspace*{-3mm}\includegraphics[width=9.3cm]{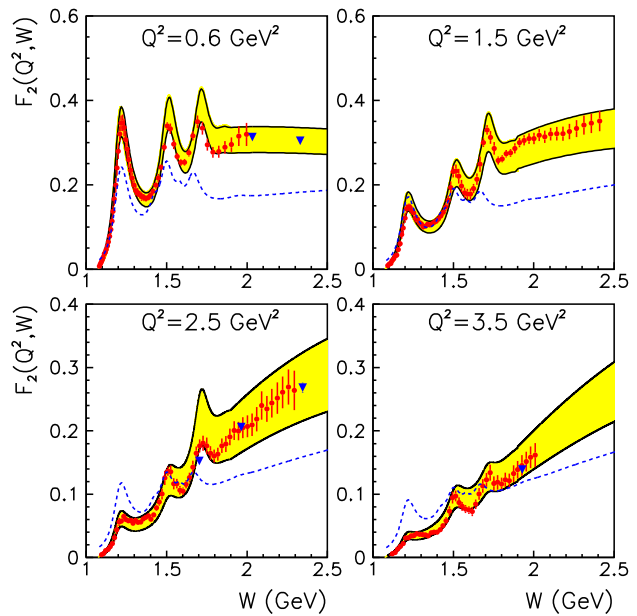}
\vspace*{-11mm}
\caption{
	Proton $F_2$ structure function versus $W$ in the resonance
	region for fixed $Q^2$. The data are from JLab (circles)
	\cite{CLAS} and SLAC (triangles) \cite{SLAC}. The shaded
	(yellow) band between the solid lines represents the uncertainty
	on our fit, while the dashed lines are obtained by summing the
	resonance fit from Ref.~\cite{Bianchi} and the nonresonant
	contributions from Ref.~\cite{Cvetic}.}
\label{fig:res}
\end{figure}

Following the analyses of Refs.~\cite{Cone,Stein,Brasse} we fit all
of the available data, including the latest from SLAC \cite{SLAC}
and JLab \cite{CLAS,HallC}, using the isobar model for each $Q^2$,
taking into account the contributions from four resonances:
$P_{33}(1232)$, $D_{13}(1520)$, $F_{15}(1680)$ and $F_{37}(1950)$.
The background contribution is taken to have the functional form 
$(1-x)^\beta/x$, which allows for a smooth transition to the 
large-$W$ region.
The fit allows the inclusive transition form factors for each of the
resonances to be constrained accurately up to $Q^2 \approx 3$~GeV$^2$.
At larger $Q^2$, where the resonance transitions are not as well
determined, we extrapolate the form factors using an exponential
form \cite{Braun}.
This introduces a relatively small uncertainty, as the resonance
contributions are strongly suppressed at large $Q^2$.

The results of our fit for the $F_2$ structure function are shown in
Fig.~\ref{fig:res} as a function of $W$ for several values of $Q^2$,
together with the uncertainties associated with the parameters of the
fit. For comparison, we also show the results obtained by adding the
contributions from resonances parametrized in Ref.~\cite{Bianchi}
and the background from Ref.~\cite{Cvetic}, which were used in the
analysis of Ref.~\cite{GH}.
The resonance parametrization \cite{Bianchi} fixes the fit parameters
from data at the real photon point, $Q^2{=}0$. To obtain transverse
and longitudinal cross sections, the $Q^2$ dependence was inferred in
Ref.~\cite{GH} using a simple {\it ansatz}.
Comparison with the data in Fig.~\ref{fig:res} shows that this
parametrization does not adequately reproduce the experimental
results in the resonance region.

In the DIS region the interference structure functions can be expressed
in terms of leading twist parton distribution functions (PDFs).
However, the range of integration in Eq.~(\ref{eq:imagi}) extends to
low $W$ and $Q^2$, beyond the region of validity of a PDF description.
To proceed we follow Ref.~\cite{GH} and approximate $F_{1,2}^{\gamma Z}$
by their electromagnetic analogs $F_{1,2}$ at very small $x$ where the
light-quark PDFs are approximately flavor independent.
Here the structure functions are proportional to a sum over products of
weak and electric charges, which for three flavors are approximately
equal \cite{GH,PDG}.

\begin{figure}[tH]
\vspace*{-6mm}
\hspace*{-3mm}\includegraphics[width=9.3cm]{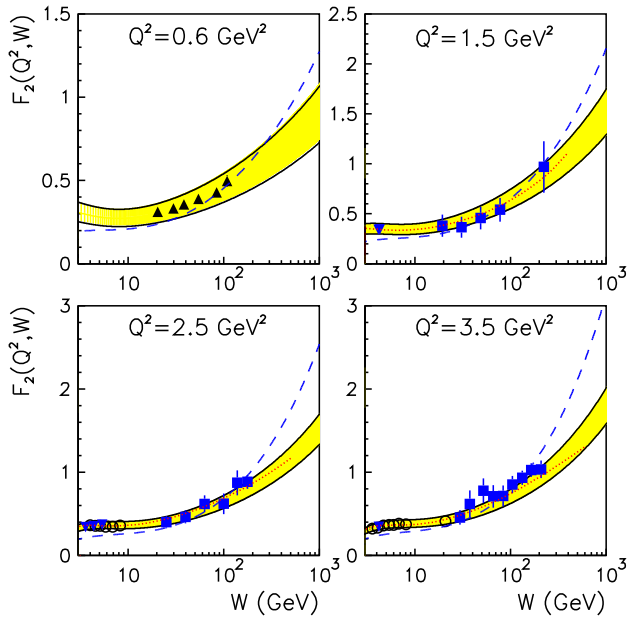}
\vspace*{-11mm}
\caption{
	As in Fig.~\ref{fig:res} but for higher values of $W$.
	The data are from SLAC (triangles) \cite{SLAC},
	NMC (circles) \cite{NMC} and H1 (squares) \cite{H1}.
	The solid lines show the fit \cite{Capella,Kaidalov} used
	in this analysis, with the (yellow) bands representing the
	uncertainty. The dashed lines are the results of the GVD
	color dipole model \cite{Cvetic}, while the dotted lines
	represent the MRST leading twist fit \cite{MRST}.}
\label{fig:dis1}
\end{figure}

For $F_2$ we use the parametrization from Refs.~\cite{Capella,Kaidalov},
which is motivated by Regge theory and valid at both high $Q^2$ and low
$Q^2$,
\begin{eqnarray}
F_2(x,Q^2)
&=& A_P\, x^{-\Delta} (1-x)^{n+4}
    \left[ \frac{Q^2}{Q^2+\Lambda_P^2} \right]^{1+\Delta}
\nonumber\\ 
&+& A_R\, x^{1-\alpha_R}(1-x)^n\!\!
    \left[ \frac{Q^2}{Q^2+\Lambda_R^2} \right]^{\alpha_R} ,
\label{eq:regge}
\end{eqnarray}
where the first term accounts for the Pomeron contribution,
while the second arises from an effective Reggeon exchange.
The parameters, and the $Q^2$ dependence of the exponents $n$ and
$\Delta$, are given given in Ref.~\cite{Capella}, with a couple
of minor adjustments to better fit the high-$Q^2$ HERA data
\cite{Derrick} and the MRST leading twist fit \cite{MRST}
({\em viz.}, the normalization of $\Delta$ increased by 8\% and
its $Q^2$ cut-off mass decreased by 5\% relative to \cite{Capella}).

\begin{figure}[tH]
\vspace*{-6mm}
\hspace*{-3mm}\includegraphics[width=9.3cm]{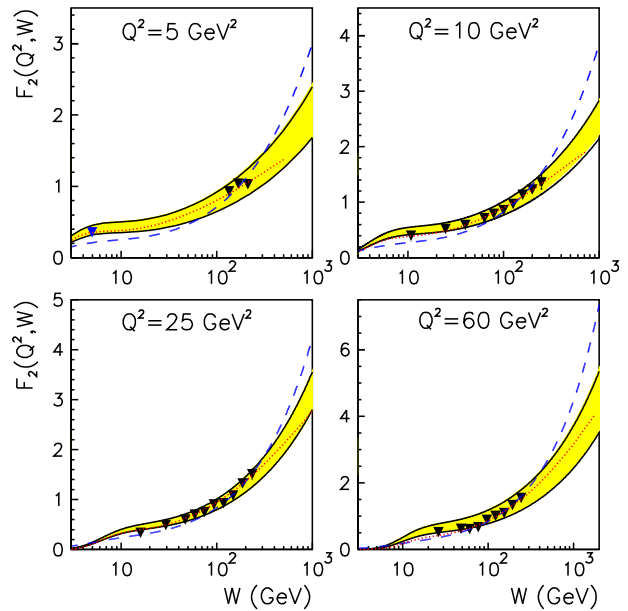}
\vspace*{-11mm}
\caption{
         As in Fig.~\ref{fig:dis1} but for
        $Q^2=5$, 10, 25 and 60~GeV$^2$.}
\label{fig:dis2}
\end{figure}

At larger $x$ ($x \gtrsim 0.4$) the flavor dependence of the PDFs
renders the interference function $\sim 30$--40\% smaller than $F_2$.
The electromagnetic structure functions therefore provide an upper
limit on $F_2^{\gamma Z}$.
However, to obtain a more accurate estimate, we use
$F_2^{\gamma Z} = (F_2^{\gamma Z}/F_2)^{\rm LT}\, F_2$, with $F_2$
given by Eq.~(\ref{eq:regge}) and the leading twist (LT) ratio
constructed from the MRST PDFs \cite{MRST}.
As Figs.~\ref{fig:dis1} and \ref{fig:dis2} demonstrate, this procedure
yields a very good description of the $W$ dependence of the available
SLAC, NMC and H1 data \cite{SLAC,NMC,H1} in the kinematics most relevant
for $\Box_{\gamma Z}^V(E)$, and also gives an excellent description
of the ZEUS data up to $Q^2 \approx 90$~GeV$^2$ \cite{ZEUS}.
The fit (\ref{eq:regge}) in the DIS region also agrees well with
the MRST parameterization \cite{MRST} of $F_2$.
In contrast, the generalized vector dominance (GVD) color dipole model
\cite{Cvetic}, used in the calculation of Ref.~\cite{GH}, slightly
underestimates the data at lower $W$, but exceeds the other fits
above $W \sim 100$~GeV.

The contribution from the $F_1$ structure function to
$\Im m\, \Box_{\gamma Z}^V(E)$ is obtained from $F_2$ and the ratio
$R = \sigma_L/\sigma_T = (1+4M^2x^2/Q^2) F_2/(2 x F_1) - 1$ of the
longitudinal and transverse cross sections. The latter has been
measured only over a limited range of $x$ and $Q^2$; however, its
contribution is numerically small, especially at large $Q^2$ and $W$.
We use the parameterization
$R = c_1 Q^2 (\exp({-c_2 Q^2}) + c_3\, \exp({-c_4 Q^2}))$, with
$c_{1\cdots 4} = \{0.014,0.07,41,0.8\}$, which provides a good
description of available data and has the correct photoproduction
and high-energy limits.
This parametrization compares favorably with the earlier SLAC fit
\cite{Rslac}, which included a mild $x$ dependence, but was restricted
to $Q^2 \gtrsim 0.3$~GeV$^2$.

\begin{figure}[bH]
\vspace*{-4mm}
\hspace*{-2mm}\includegraphics[width=7.0cm]{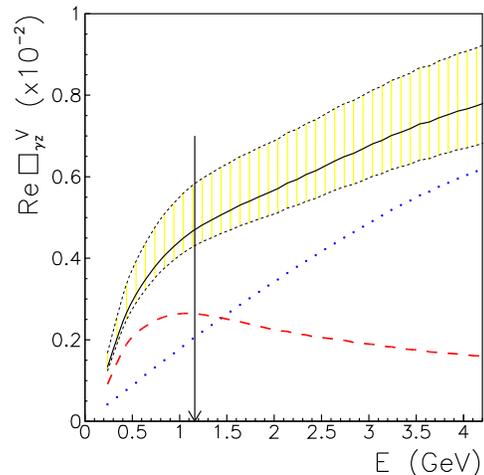}
\vspace*{-5mm}
\caption{
	$\gamma Z$ box correction $\Box_{\gamma Z}^V(E)$ to $Q_W^p$
	as a function of electron energy $E$, showing the resonant
	(dashed) and nonresonant (dotted) contributions, as well as
	the sum (solid) and the overall (asymmetric) uncertainty
	(shaded). The vertical arrow at $E=1.165$~GeV indicates
	the energy of the ${\rm Q_{weak}}$ experiment.}
\vspace*{-4mm}
\label{fig:del}
\end{figure}

Performing the dispersion integration in Eq.~(\ref{eq:DR}), the
result for $\Re e\, \Box_{\gamma Z}^V(E)$ is shown in Fig.~\ref{fig:del}
as a function of the incident electron energy $E$.
Although the integration in principle involves an infinite range of
$W$ and $Q^2$, in practice we find that around 80\% of the value of
$\Re e\, \Box_{\gamma Z}^V(E)$ at the energy relevant to ${\rm Q_{weak}}$ 
comes from energies below 4~GeV, where the $Q^2$ range extends to 
$\sim 6$~GeV$^2$, and $W$ to $\sim 3$~GeV.
This is fortunate as it is precisely in this region that a wealth
of very accurate data exists from JLab \cite{CLAS,HallC}.

In Fig.~\ref{fig:del} the nonresonant contribution is small at
low energies, but rises logarithmically with increasing $E$.
The resonance contribution increases steeply to a maximum
at $E \sim 1$~GeV, before falling off like $1/E$.
The resonant and nonresonant contributions to
$\Re e\, \Box_{\gamma Z}^V(E)$ are 0.0026 and 0.0021, respectively, at
the energy relevant for the ${\rm Q_{weak}}$ experiment, $E=1.165$~GeV.
We should note, however, that this separation is somewhat arbitrary,
as only the physically meaningful, total cross section enters into
our fit.

The combined correction to $Q_W^p$ at the ${\rm Q_{weak}}$ energy
is then $0.0047^{+0.0011}_{-0.0004}$, or $6.6^{+1.5}_{-0.6}\, \%$
of the Standard Model value $0.0713(8)$ for $Q_W^p$, with the error
band obtained from the uncertainty in the fit parameters using a
variational method.
In comparison, the correction found in Ref.~\cite{GH} was $\approx 0.003$.
The major difference with our value arises from the additional factor 2 
in Eq.~(\ref{eq:imagi}), which has been verified by the independent 
checks discussed earlier, as well as our use of more recent
electroproduction data and the correct relation between structure
functions and the virtual photon cross section.
The correction is important for the interpretation of the ${\rm Q_{weak}}$
experiment, given its projected uncertainty of $\pm 0.003$ \cite{QWEAK}.
It is also critical to the physical interpretation of the experiment
which is expected to constrain possible sources of parity violation
from beyond the Standard Model at a mass scale of $\gtrsim 2$~TeV 
\cite{YoungSM}.
While the uncertainty in the result reported here is satisfactory
from the point of view of ${\rm Q_{weak}}$, it can be further reduced
by incorporating the new inclusive parity-violating data in the
resonance region which should be taken soon at JLab \cite{PVDIS}.

\begin{acknowledgments}

We thank R.~Carlini, M.~Gorchtein, D.~Schildknecht and W.~van~Oers
for helpful discussions and communications.
This work was supported by the DOE contract No. DE-AC05-06OR23177,
under which Jefferson Science Associates, LLC operates Jefferson Lab,
NSERC (Canada), and the Australian Research Council through an 
Australian Laureate Fellowship (AWT).

\end{acknowledgments}


\end{document}